# Induced Quantum Entanglement of Nuclear Metastable States of $^{115}$In

D. L. Van Gent, Nuclear Science Center, Louisiana State University, Baton Rouge, USA

**Abstract**

Experiments conducted in our laboratory conclusively demonstrated that at least 20% of $^{115}$In metastable states become quantum entangled (QE) during gamma photo-excitation processes where a significant fraction of the photo-excitation gamma (E > 1.02 MeV) are QE. In addition, it was found that the half-life of $^{115m}$In populations in identical photo-excited indium foils varied as much as 70% depending on whether the 99.999% purity indium foils were photo-excited with a High Intensity $^{60}$Co Source (HICS) or a Varian CLINAC (Compact Linear Accelerator) with average energy 2 MeV and maximum energy 6 MeV Bremsstrahlung photo-excitation quanta. Decay kinetics of $^{115m}$In populations in indium foils demonstrate that these metastable states are primarily QE in pairs when photo-excited in the HICS apparatus and at higher orders of entanglement of triplets and possibly quadruplets when photo-excited with the CLINAC. It appears that QE gamma photons can transfer quantum entangled properties to radioactive metastable states.

## 1. Introduction

The possibility of quantum entanglement over macroscopic distances was first alluded to by Einstein, Podolsky, and Rosen [1]. They wrote with strong conviction that General Relativity and QED are fundamentally at odds with each other in this respect, since QED seems to indicate the possibility of "instantaneous" transfer of quantum information over long distances. It is widely acknowledged that, in theory, quantum noise can be transferred instantaneously to any point in the universe via QE systems.

We report here for the first time strong evidence that quantum entanglement of gamma photons can be transferred for extended periods of time into nuclear radioactive metastable states of certain photo-excited metals. Paired QE nucleonic metastable states must conform to quantum spin and angular momentum conservation laws even when separated by macroscopic distances similar to QE paired photons.

## 2. Methodology

It is well known that low energy photon pairs from atomic radiative cascade are entangled [2,3]. In the experiments reported here, entangled gamma photons were produced from both radio-isotopic $^{60}$Co nuclear decay and from a CLINAC Compact Linear Accelerator. Stable indium ($^{115}$In) in 99.999% pure natural indium foils were photo-excited into ordinary metastable states ($^{115m}$In) and QE metastable states.

Figure 1 depicts the decay scheme of $^{115m}$In.

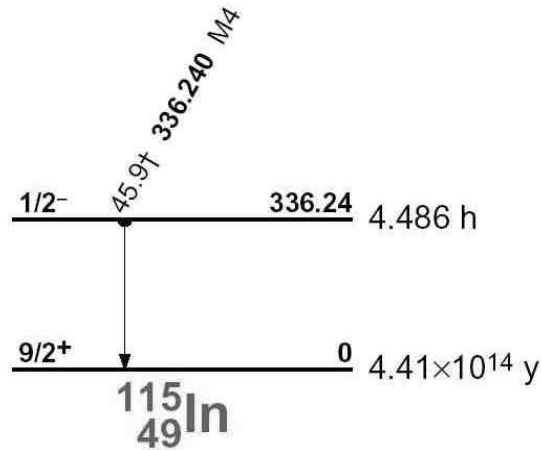

Figure 1. Decay scheme of $^{115m}$In from Table of Isotopes, CD-ROM, 8$^{th}$ edition, Version 1.0, Richard B. Firestone, Laurence Berkeley National Laboratory, University of California.

Figure 2 depicts the decay scheme of $^{60}$Co with transition from $^{60}$Co to $^{60}$Ni.

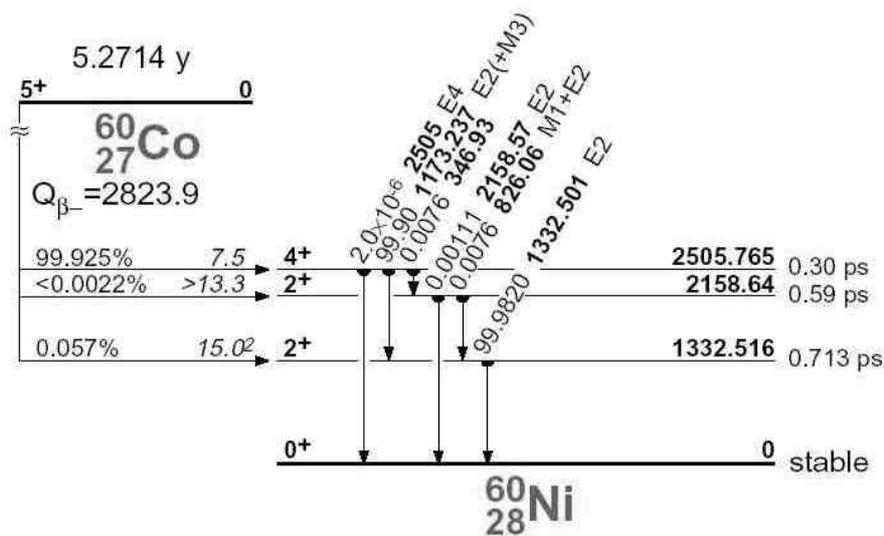

Figure 2. Decay scheme of $^{60}$Co from Table of Isotopes, CD-ROM, 8$^{th}$ edition, Version 1.0, Richard B. Firestone, Laurence Berkeley National Laboratory, University of California.

The schematic indicates that 99.90% of $^{60}$Co decays via beta emission to energy level 2505.7 keV and 99.9% decays from that energy level to 1332.5 keV with an emission of a 1173.2 keV gamma photon, followed on the average of 0.7 picoseconds later by another gamma photon of 1332.5 keV which drops the nucleus to ground state of stable $^{60}$Ni. This is a typical example of a cascade nuclear emission of two sequential gamma photons with slightly differing energies that are QE.

The HICS utilized for QE photo-excitation of indium foils is a Shepherd Upright Irradiator, Model 484R which contains 111,000 GBq (3000 Ci) of $^{60}$Co. The gamma photon flux at the indium foil position is approximately $10^{14}$ photons cm$^{-2}$ sec$^{-1}$. Figure 3 depicts the inner irradiation chamber and where the foils were taped.

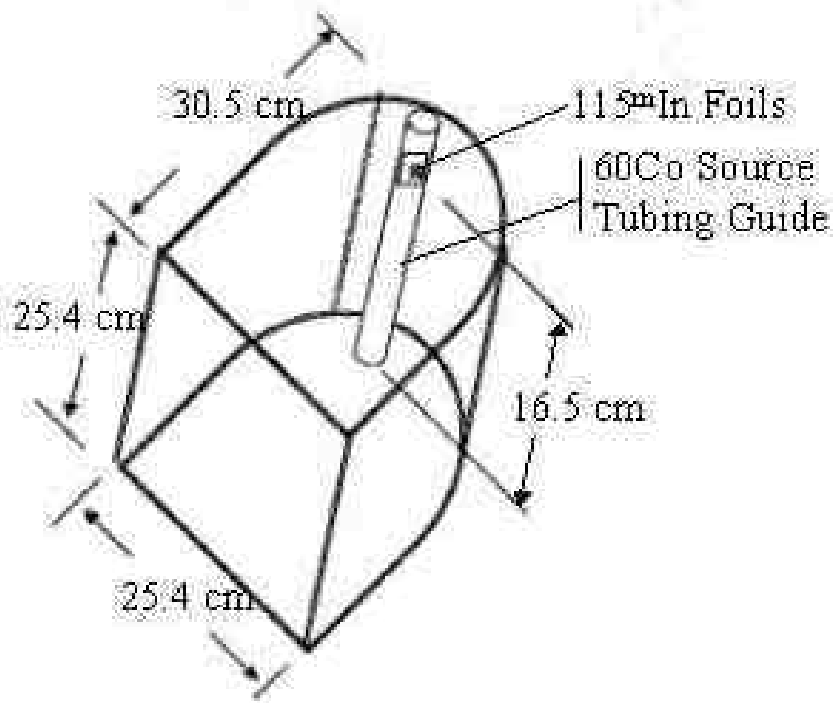

Figure 3. Diagram of the irradiation chamber of the Shepherd 484R Gamma Irradiator HICS (Courtesy of Dr. Wei-Hsung Wang, 2004).

The Compact Linear Accelerator (CLINAC) produces high energy QE photons via Bremsstrahlung production. The CLINAC linear accelerator accelerates electrons to energies up to a maximum of 6 MeV before the electrons interact with a tungsten target, producing Bremsstrahlung cascade of photons and QE photons. Figure 4 is a simple schematic of a CLINAC and Figure 5 is a typical photon Bremsstrahlung spectrum produced by a CLINAC. A comprehensive review and characterization of the CLINAC is available in the literature [4,5].

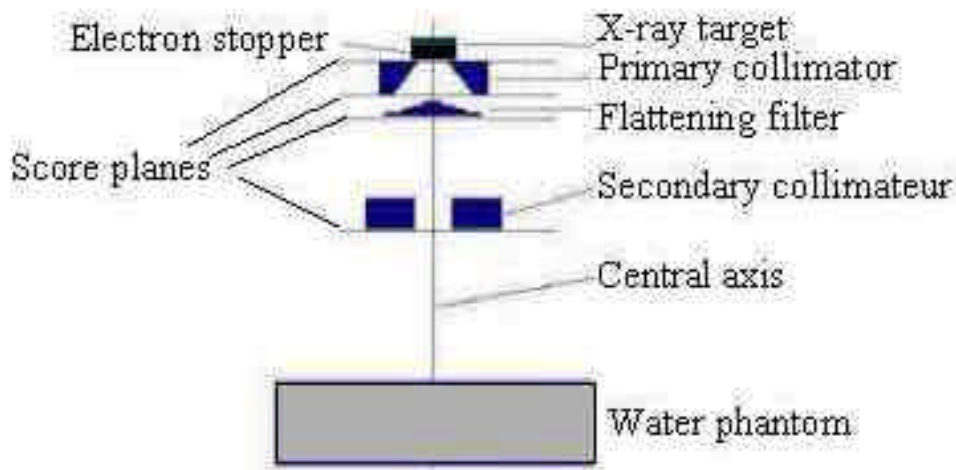

Figure 4. CLINAC Schematic (Courtesy of Sameer S. A. Natto, 1992)

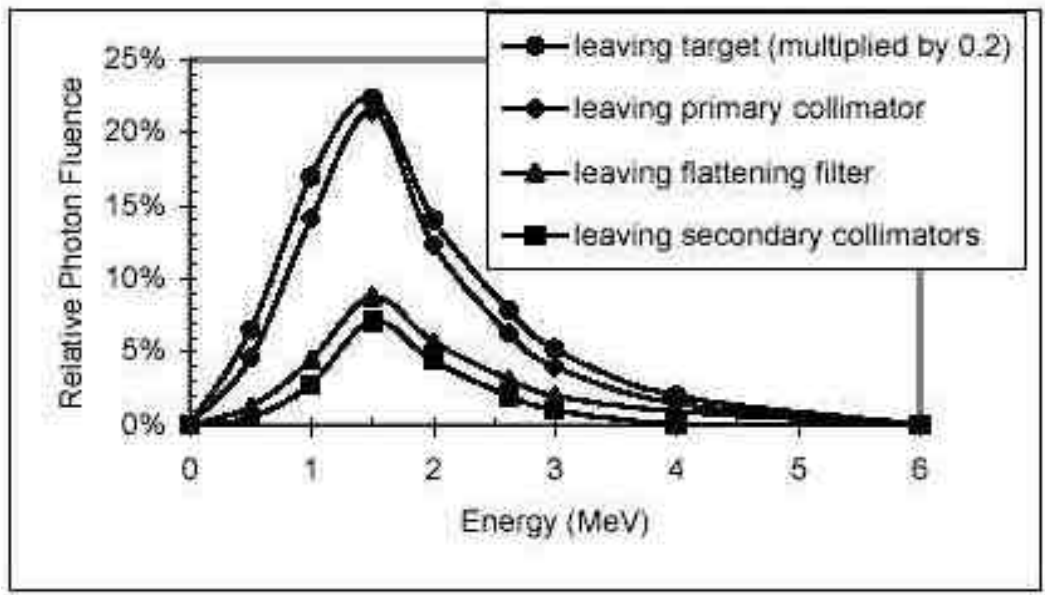

Figure 5. Typical CLINAC Bremsstrahlung spectrum (Courtesy of Sameer S. A. Natto, 1992)

Several nuclear metastable states can be artificially produced via photo-excitation by high-energy QE photons from the sources described above. Table 1 shows examples of selected artificially produced metastable nuclear isotopes. Indium foils were chosen for this study since indium is commercially available in ultra high purity form (99.999%) and the photo-excitation cross-section is quite large. $^{115m}$In has a half-life of almost exactly 268 minutes and thus is useful for rapid short-term experimentation with excitation saturation occurring in less than 48 hours per indium foil irradiation.

| Nucleus | Symbol | Abundance % | Half-Life | Gamma keV |
|---|---|---|---|---|
| Niobium | 93Nb41 | 100 | 16.3 y | 31.8 |
| Cadmium | 111Cd48 | 12.8 | 48.54 m | 396.2 |
| Cadmium | 113Cd48 | 12.2 | 14.1 y | 263.5 |
| Indium | 115In49 | 95.7 | 4.48 h | 336.2 |
| Tin | 117Sn50 | 7.7 | 13.6 y | 314.6 |
| Tin | 119Sn50 | 8.6 | 293 d | 60.5 |
| Tellurium | 125Te52 | 7.1 | 57.4 d | 144.8 |
| Xenon | 129Xe54 | 26.5 | 8.8 d | 238.1 |
| Xenon | 131Xe54 | 21.2 | 11.8 d | 163.9 |
| Hafnium | 178Hf72 | 27.3 | 31 y | 574/…./93 |
| Hafnium | 179Hf72 | 13.6 | 25 d | 453/…./122 |
| Iridium | 193Ir77 | 62.7 | 10.5 d | 80.2 |
| Platinum | 195Pt78 | 33.8 | 4 d | 259.3 |

Table 1.  Some Metastable Nuclei with stable ground state

After the foils were irradiated, they were counted for nuclear gamma emission at the characteristic energy of 336 keV for $^{115m}$In on a Canberra high purity intrinsic germanium gamma counting system with an Ortec Low-Level Background lead, copper, and steel shield for one minute for each consecutive counting period. Each spectral file was automatically saved on hardware and each indium foil gamma counting session continued repeatedly in the same manner utilizing macro language for the Multichannel Analyzer software package.

**3.  Results**

Typical counting results for a 30 hour HICS irradiated Indium foil are depicted in Figure 6 and the counting results for CLINAC 30 minute irradiated indium foil are depicted in Figure 7.

There was a marked difference between observed half-lives of $^{60}$Co irradiated foil versus CLINAC irradiated foil. The $^{60}$Co irradiated foil half-life was 251 minutes for the first 100 minutes of counting with $r^2$ = 0.86 with error of y-estimate 2.77%. The CLINAC irradiated foil half-life for the first 100 minutes of counting was 184 minutes with $r^2$ = 0.48 with error of y-estimate 11.6%. The standard neutron activated half-life determination for $^{115m}$In is 268 minutes [5].  These results are typical of over 10 replicates for the $^{60}$Co irradiated foil irradiations and 4 replicates of the CLINAC foil irradiations

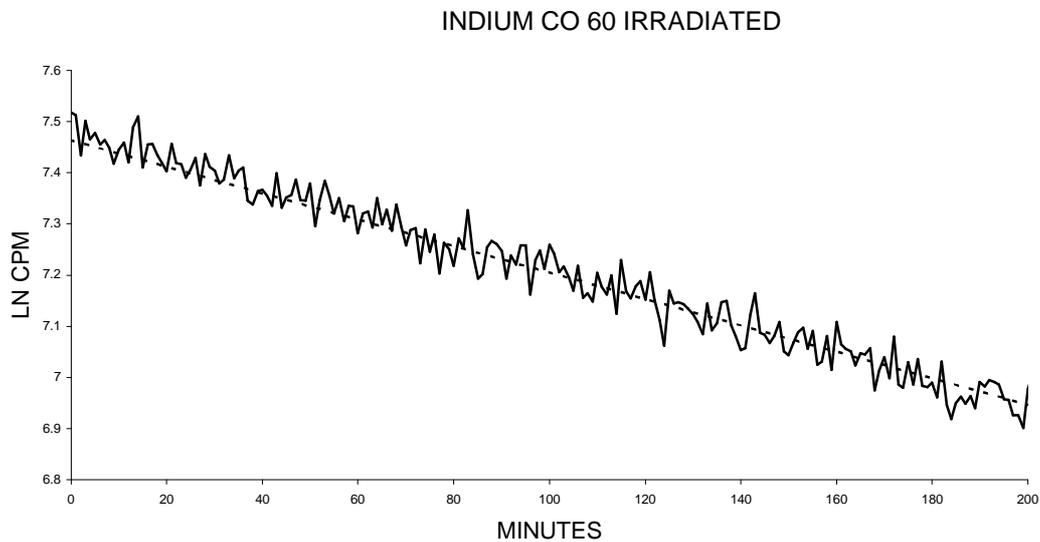

Figure 6. Graph of natural log of $^{115m}$In gamma cpm vs. minutes for HICS irradiated indium foil. The dashed line is a linear fit of the data. There is a slight non-linearity of the data.

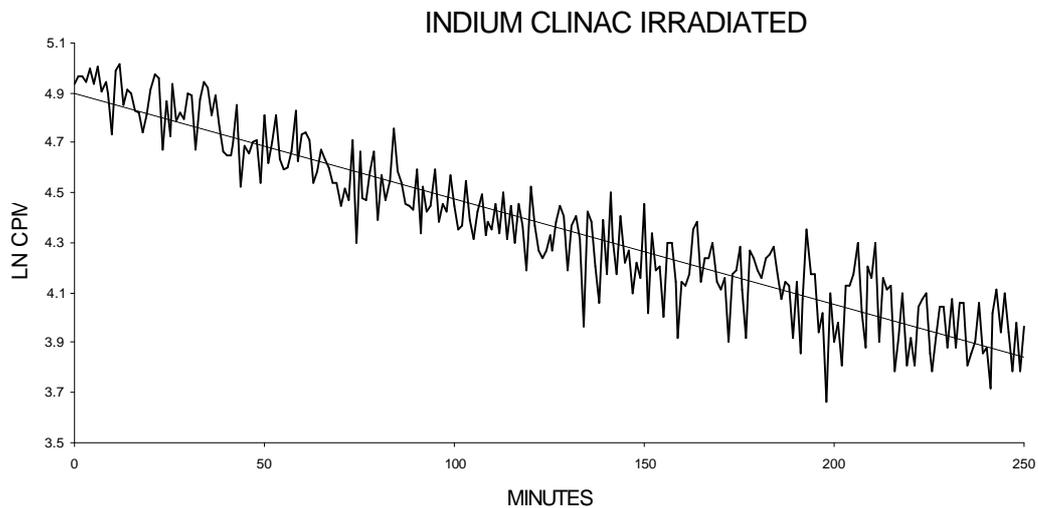

Figure 7. Graph of natural log of $^{115m}$In gamma cpm vs. minutes for CLINAC irradiated indium foil. The dashed line is a linear fit of the data. The non-linearity of the data is obvious.

A numerical exponential finite time element model was fitted to the data in both cases. For HICS photo-excited $^{115m}$In foil populations, the model best fit indicated that 3.5% of metastable nuclei are QE and are entirely doublets. For CLINAC photo-excited $^{115m}$In foil populations, the model best fit indicated that 9% of metastable nuclei are QE as doublets and 9% are QE as triplets, since during the cascade Bremsstrahlung one electron can produce three or four entangled gamma with an energy suitable for Indium 115 excitation.

## 4. Conclusion

The data from this experiment strongly indicates that high energy gamma and Bremsstrahlung photons are entangled to various degrees. The rationale for shorter half-lives of entangled metastable states is quite straight forward since one would expect that when two metastable states are entangled, the decay probability should double from 0.002586/min to 0.005172/min and so on for triplets, quadruplets, etc. These different probabilities are easily deconvoluted from compound exponential data with QE metastable state decay.

This photon quantum entanglement can be transferred to metastable nuclei of photo-excitable elemental metals. The data reported here demonstrate that there are multiorders of metastable state entanglement possible, depending upon the maximum photon energy available. Since the HICS source yields gamma photons at approximately 1100 and 1300 keV, only doublets can be formed in that case. On the other hand, the CLINAC yields a maximum energy of 6 MeV photons which could theoretically create up to quintuplet QE nuclei. During the analyses of the experimental data via numerical modeling, doublets and triplets were easily observable and necessary for the constraints of the model.

**Acknowledgement**

I thank Professor Robert Desbrandes (LSU emeritus from Petroleum Engineering) for his help in the experiments and their interpretation. I also thank the Veterinary School of LSU for the use of its CLINAC accelerator and the Nuclear Science Center of LSU for using its HICS Cobalt 60 source and Germanium gamma counter.